\documentclass[9pt,conference]{IEEEtran}
\usepackage[preprint]{waspaa25}
\usepackage{bm} % for bold math symbols (incl. Greek letters) with \bm{}
\usepackage{xcolor}
\usepackage{graphicx}
\usepackage{booktabs}
\usepackage{float}

\title{TACOS \raisebox{-0.2em}{\includegraphics[height=1.1em]{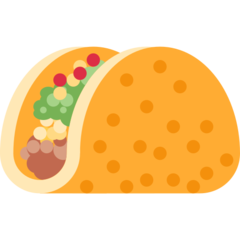}}: Temporally-aligned Audio CaptiOnS \\ for Language-Audio Pretraining}

\name{Paul Primus$^1$, Florian Schmid$^1$, and Gerhard Widmer$^{1,2}$\thanks{The LIT AI Lab is supported by the Federal State of Upper Austria. Gerhard Widmer's work is supported by the European Research Council (ERC) under the European Union's Horizon 2020 research and innovation programme, grant agreement No 101019375 (Whither Music?). We acknowledge the support of Label Studio (https://labelstud.io) for providing access to their Enterprise version through their Academic Program.}}
\address{
$^1$Institute of Computational Perception (CP-JKU),  
$^2$LIT Artificial Intelligence Lab\\
Johannes Kepler University, Austria
}

\begin{document}

\maketitle

\begin{abstract}
Learning to associate audio with textual descriptions is valuable for a range of tasks, including pretraining, zero-shot classification, audio retrieval, audio captioning, and text-conditioned audio generation.
Existing contrastive language-audio pretrained models are typically trained using global, clip-level descriptions, which provide only weak temporal supervision.
We hypothesize that CLAP-like language-audio models -- particularly, if they are expected to produce frame-level embeddings -- can benefit from a stronger temporal supervision.
To confirm our hypothesis, we curate a novel dataset of approximately 12,000 audio recordings from Freesound, each annotated with single-sentence free-text descriptions linked to a specific temporal segment in an audio recording.
We use large language models to clean these annotations by removing references to non-audible events, transcribed speech, typos, and annotator language bias.
We further propose a frame-wise contrastive training strategy that learns to align text descriptions with temporal regions in an audio recording and demonstrate that our model has better temporal text-audio alignment abilities compared to models trained only on global captions when evaluated on the AudioSet Strong benchmark.
The dataset and our source code are available on Zenodo and GitHub, respectively.
\end{abstract}

\section{Introduction}
\label{sec:intro}

In recent years, the acoustic signal processing community has made significant progress in audio classification, advancing tasks such as Audio Tagging~\cite{Gong21AST, Chen22HTSAT,Koutini22Patchout,Schmid23Efficient,Liu23CAT,Schmid24DCNN}, Sound Event Detection (SED)~\cite{Nam22FDC,Shao24ATST,Li24SSTST,Cai24MATSED}, and Acoustic Scene Classification~\cite{Kim22DG,Schmid23CPMobile,Han25DataEfficient}. These systems have become increasingly performant, particularly due to the availability of large-scale datasets such as AudioSet~\cite{Gemmeke17AudioSet,Hershey21StrongLabels}. However, they remain constrained to predicting a fixed set of predefined labels.

To overcome this limitation, a growing line of research has begun exploring more flexible audio-language models that align natural language text with audio, aiming for a richer and more comprehensive understanding of acoustic content. Multimodal audio-language learning tasks include text-based audio retrieval~\cite{Koepke23Benchmark,Mei22MetricLearning,Primus23Retrieval,Lou22Context}, audio captioning~\cite{Mei22AACOverview,Drossos17AAC,Gontier21BART,Mei21ACT}, text-conditioned audio generation~\cite{Kreuk23AudioGen,Huang23MakeAnAudio,Liu23AudioLDM,Borsos23AudioLM}, and audio question answering~\cite{Gong24LTU,Lipping22ClothoAQA,Fayek20AQA}. Training models for these tasks requires datasets that pair audio recordings with corresponding textual descriptions—so-called \textit{audio captions}. Several widely used datasets, including Clotho~\cite{Drossos20Clotho}, AudioCaps~\cite{Kim19AudioCaps}, MACS~\cite{Morato21MACS}, and WavCaps~\cite{Mei24WavCaps}, have driven progress in these areas. Despite differences in scale and annotation methodology, they share a common characteristic: they provide \textit{global} captions that describe an entire audio clip as a whole. We refer to such annotations as \textit{weak} captions, as they do not specify when particular events occur within the clip.

Despite recent progress, Wu et al.~\cite{Wu23NaturalLanguage} have highlighted key limitations in current audio-text models: notably, their insensitivity to the temporal ordering of events described in captions. They identify two primary causes for this behavior: (1) the use of global temporal pooling in CLAP-like~\cite{Elizalde23CLAP} architectures, which collapses temporal structure into a single representation, and (2) the lack of training data containing captions that express complex temporal relationships, such as event simultaneity and ordering. 

To address these limitations, we introduce the concept of \textit{strong} captions—free-text descriptions that are temporally aligned with specific segments of an audio clip, defined by onsets and offsets. Rather than collapsing the entire audio into a single global representation, this setup enables a frame-wise extension to the contrastive learning used in CLAP-style~\cite{Elizalde23CLAP} models, which aligns temporally-localized audio segments (e.g., audio embeddings at a resolution of 20 ms) with their corresponding textual descriptions. We hypothesize that the temporally fine-grained supervision enables models to capture the temporal structure and relationships between sound events more effectively.

This paper makes the following contributions:

\begin{itemize}
\item We release \textit{TACOS}\footnote{Dataset: https://zenodo.org/records/15379789}, a dataset of 12,358 audio recordings annotated with 47,748 \textit{strong} audio captions, along with a \textit{weak} caption per file that is automatically derived from the strong ones using ChatGPT (Section~\ref{sec:tacos}).
\item We propose a novel \textit{frame-wise} contrastive training approach that aligns temporally-localized audio segments with corresponding textual descriptions (Section~\ref{sec:method}).
\item We demonstrate that our method learns meaningful local audio-text correspondences by evaluating models trained on \textit{TACOS} in a text-based Sound Event Detection setting using AudioSet Strong~\cite{Hershey21StrongLabels} (Sections~\ref{sec:experiments} and \ref{sec:results}).
Our implementation is available on GitHub\footnote{\label{note1}Source code and details: https://github.com/OptimusPrimus/tacos}.
\end{itemize}

\section{Creation of the Tacos Dataset}
\label{sec:tacos}

\subsection{Audio Data Collection}

We collect audio from the Freesound platform~\cite{Font13Freesound} via its public API, guided by a two-level ontology comprising seven superclasses (e.g., \textit{Human Sounds}, \textit{Animals}, \textit{Machinery/Industrial}) and 59 fine-grained subclasses (e.g., \textit{Animals} $\rightarrow$ \textit{Rooster Crow}). For each subclass, we define semantically relevant search terms (e.g., {"rooster crow", "cock-a-doodle-doo", "chicken crow"} for \textit{Rooster Crow}) to retrieve clips\footnotemark[2]. We download up to 300 top-ranked MP3 files per term, applying filters to ensure a minimum sampling rate of 32~kHz, a bit depth of 16 bits, and a duration between 15 and 300 seconds. This yields 19,230 clips, averaging 325.93 files per subclass with \textit{Speech} as the most populous (854 clips) and \textit{Hiccup} the least (45 clips).

We discard 120 clips flagged as explicit or containing blacklisted terms such as profanity or slurs. To focus on real-world recordings, we remove 3,208 synthetic clips based on keywords such as “synth”, “generated”, or “artificial”. Most of these come from the \textit{Musical Instruments} superclass, mainly due to artificially rendered instruments.

\begin{table*}[!t]
\centering
\caption{Annotated regions for a sample file (\textit{667739.mp3}) labeled by two annotators (A and B), showing original and cleaned captions. The weak caption is generated automatically from the individual region descriptions by employing OpenAI’s \textit{gpt-4o-mini-2024-07-18} model~\cite{openai2024gpt4omini}.}
\label{tab:example-two-annotators}
\scriptsize
\begin{tabular}{cllp{.38\textwidth}p{.38\textwidth}}
\toprule
\textbf{Annotator} & \textbf{Onset (s)} & \textbf{Offset (s)} & \textbf{Original Caption} & \textbf{Cleaned Caption} \\
\midrule
A & 0.000  & 2.605  & Train approaching with horn. & A train approaches, sounding its horn. \\
A & 2.624  & 20.848 & Train going by. & A train passes by. \\
B & 0.040  & 1.746  & A train horn blares in the distance. & A train horn blares in the distance. \\
B & 1.760  & 2.969  & A train drives by at a deafening volume and close distance. & A train passes by at a deafening volume and close distance. \\
B & 2.982  & 20.848 & A train drives off into the distance gradually decreasing in volume. & A train moves away, gradually decreasing in volume. \\
\midrule
\multicolumn{5}{p{\textwidth}}{\textbf{Generated weak caption:} A train approaches sounding its horn and then passes by.} \\
\bottomrule
\end{tabular}
\end{table*}

\subsection{Audio Processing}
\label{subsec:audio_processing}

Our audio preprocessing is inspired by the procedures used for the Clotho dataset~\cite{Drossos20Clotho}. We normalize waveforms to the range \([-1, 1]\), and trim leading and trailing silence based on an energy threshold of 60~dB below the maximum amplitude. We discard clips shorter than 15~s post-trimming and resample the remaining audios to 32~kHz.

To ensure consistency in clip duration, we randomly sample a target duration between 15 and 30 seconds and extract a segment of that length from clips longer than 30 seconds. To prioritize dynamic and information-rich content, we select the segment with the highest total energy. To avoid abrupt edges at segment boundaries, we apply a 16~ms Hamming window to the start and end of the extracted audio.

\subsection{Collecting Temporally-Aligned Captions}

The dataset is annotated by 337 students as part of a university machine learning course, with each student labeling 40 randomly selected clips from the 15{,}642 remaining audio files.

Annotations are collected using the Label Studio~\cite{labelstudio} Enterprise platform. The annotation interface displays waveforms, allowing annotators to mark temporal regions and assign free-text captions to them. A time-aligned spectrogram is shown alongside the waveform to support precise identification of onsets and offsets. Additionally, annotators have access to Freesound metadata (tags, descriptions). While this introduces potential bias—annotators might rely more on textual cues than auditory perception, as discussed in~\cite{Drossos20Clotho}—we consider this trade-off acceptable given the annotators' non-native English background. In practice, the metadata reduces perceptual ambiguity and supports more consistent, accurate labeling.

Annotators receive detailed guidelines to ensure consistent and high-quality annotations. Key instructions include: (1) create a separate annotation region for each distinct sound source perceived; (2) split repeated sounds of the same source if a perceivable pause occurs; and (3) write captions that are independent of each other. Each caption must be a single English sentence structured around four dimensions: \textit{Source and Action} (what produces the sound?), \textit{Descriptor} (what does it sound like?), \textit{Temporal} (does the sound evolve over time?), and \textit{Context} (where or in what environment does it occur?). The guiding principle given to annotators is: \textit{if all labeled regions are removed, the clip should be silent}. To assess annotation quality, around 10\% of files are labeled by multiple annotators to measure consistency and adherence to guidelines.

\subsection{Postprocessing Captions}

Although most annotators provided clear, well-structured sentences, captions often contain grammatical, spelling, or punctuation errors due to the annotators’ sloppiness. To improve quality, we automatically refine captions using OpenAI’s \textit{gpt-4o-mini-2024-07-18} model~\cite{openai2024gpt4omini}, guided by a carefully crafted prompt. In addition to correcting language issues, the model is instructed to remove transcribed speech, named entities, and references to other captions while preserving meaning and structure. 

We also use the same model to generate weak captions—single-sentence summaries of entire clips—by merging the temporally aligned captions. These summaries are constrained to about 20 words, limited to the provided content, and styled consistently with datasets like AudioCaps~\cite{Kim19AudioCaps} and Clotho~\cite{Drossos20Clotho}. This enables us to assess the quality of our dataset beyond strong captions by training a standard text-based audio retrieval model on the generated weak captions and evaluating its performance on Clotho, as discussed in Section~\ref{sec:results}.

Table~\ref{tab:example-two-annotators} shows original and cleaned captions from two annotators, along with the generated weak caption. While their timing and phrasing differ slightly, both describe a train horn followed by the train passing. Cleaned captions enhance fluency—for example, “A train drives off into the distance” becomes “A train moves away”—though many well-formed captions remain unchanged, such as “A train horn blares in the distance.” The weak caption concisely summarizes the strong captions while preserving event order: it accurately reflects that the horn sounds as the train approaches, followed by the train passing by.

\begin{figure}[!t]
\centering
\includegraphics[width=\linewidth]{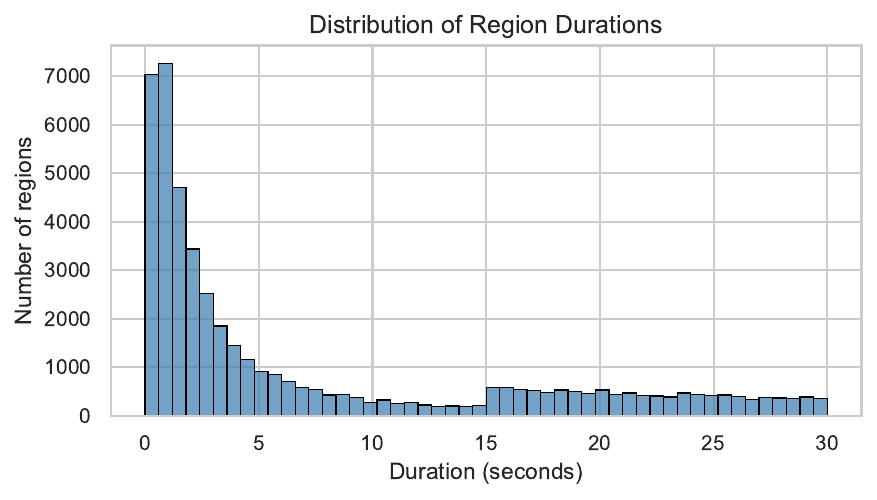}
\vspace{-0.5cm} % GW: sorry fuer den Hack ..
\caption{Distribution of annotated region durations.}
\label{fig:region-duration-hist}
\end{figure}

\subsection{Annotation Statistics and Dataset Splits}

A total of 12,358 audio recordings are annotated, including 1,031 with duplicate annotations from different annotators. In total, 47,748 annotated regions are collected, averaging 3.57 regions per clip. The dataset spans 76.59 hours of audio, with annotated regions totaling 97.99 hours due to overlaps. After merging overlapping intervals, the average temporal coverage per clip is 92.84\%. Captions average 6.89 words, with a standard deviation of 3.52. The caption vocabulary comprises 5,952 unique words without lemmatization or stop-word removal, and 5,106 with both applied. For comparison, \cite{Morato21MACS} reports 4,373 for Clotho~\cite{Drossos20Clotho}, 2,775 for MACS~\cite{Morato21MACS}, and 5,218 for AudioCaps~\cite{Kim19AudioCaps}. Despite AudioCaps being much larger (57,188 clips), TACOS shows comparable or greater lexical diversity—likely due to our ontology-based collection strategy, which promotes diverse audio content and descriptions.

Figure~\ref{fig:region-duration-hist} illustrates the distribution of annotated region durations. Many regions are short, indicating that numerous sound sources are only briefly audible within a clip. This underscores a key advantage of strong captions over weak ones: rather than labeling the entire clip, strong captions precisely localize short events (e.g., “someone sneezes”) in time. This temporal precision may benefit contrastive learning, where brief events could otherwise be diluted by global pooling in CLAP-style models. The noticeable spike at 15 seconds corresponds to the minimal clip length (15–30 seconds) and the presence of annotations spanning the full duration of some files, reflecting long or continuous sound events.

We define fixed train/test splits for TACOS, with 10,358 files for training and 2,000 reserved for testing. The split preserves the subclass distribution from the ontology (Section~\ref{subsec:audio_processing}), ensuring that both splits remain diverse and representative of the full ontology.

\section{Contrastive Training with Temporally-Strong Free-text Annotations}
\label{sec:method}

Following prior work on audio-text models \cite{Elizalde23CLAP, Drossos20Clotho, Wu23LION_CLAP}, we adopt a dual-encoder architecture to project audio and text into a shared embedding space using two modality-specific encoders. Unlike existing approaches that produce a single global representation for an entire audio recording, our model generates a sequence of frame-level embeddings that capture temporally-localized audio semantics.

To train our model, we introduce a frame-level contrastive learning objective that leverages the temporal onsets and offsets associated with free-text annotations in TACOS. This enables fine-grained alignment between arbitrary text descriptions and frame-level audio embeddings, offering greater flexibility than conventional CLAP-style models \cite{Elizalde23CLAP} or supervised sound event detection systems.
%An overview of the proposed training procedure is shown in Figure~\ref{TODO}. (no space...)

Let $\mathit{audio}^i$ denote an audio recording, and let $r^i = \{(\mathit{onset}^i_j, \mathit{offset}^i_j, \mathit{text}^i_j)\}_{j=1}^{M}$ represent its set of annotated regions with corresponding temporal boundaries and free-text descriptions. During training, we sample a batch of $N$ audio recordings and their region sets, denoted as $\mathcal{B} = \{(\mathit{audio}^i, r^i)\}_{i=1}^{N}$. Each audio recording $\mathit{audio}^i$ and its associated descriptions $\mathit{text}^{i}_j$ are independently encoded by an audio encoder $f_a$ and a text encoder $f_t$:

\begin{equation}
A^i = f_a(\mathit{audio}^i) \in \mathbb{R}^{T \times D}, \quad d^{ij} = f_t(\mathit{text}^{i}_j) \in \mathbb{R}^{D}
\end{equation}
Here, $D$ is the dimensionality of the shared embedding space, and $T$ is the number of frame-level audio embeddings produced for a given recording. For clarity, we fix a particular batch sample $i$ and region $j$, and omit indices in the following explanation. To find all frame-embedding vectors in $A$ that correspond to region $r$, we compute the index of the first and last frame as:
\begin{equation}
t_{\text{on}} = \left\lfloor \frac{\mathit{onset}}{\delta} \right\rfloor, \quad
t_{\text{off}} = \left\lceil \frac{\mathit{offset}}{\delta} \right\rceil
\end{equation}

where $\delta$ is the duration of an audio frame. 

To derive the frame-wise contrastive loss, we focus on the individual audio embeddings in region $r$, i.e., $A_t$ for $t \in [t_{\text{on}}, t_{\text{off}}]$. Let \( d^+ \in \mathbb{R}^D \) denote the text embedding corresponding to the current frame $A_t$. The set of negative text embeddings for the fixed batch sample $i$ is defined as:
\begin{equation}
\mathcal{D}^{-} = \{d^{-} \mid d^{-} \text{ originates from a batch sample } \ne i\}
\end{equation}
That is, $\mathcal{D}^{-}$ consists of all region descriptions from other audio recordings in the batch. The set of negatives $\mathcal{D}^{-}$ is shared across all regions of batch sample $i$. The similarity between an audio embedding vector $A_t$ and a text embedding vector $d \in \{d^+, d^-\}$ is computed using a scaled dot product:
\begin{equation}
s_t(d) = \frac{A_t^\top d}{\tau}
\end{equation}
where \( \tau > 0\) is a hyper parameter which re-scales the similarity scores.
We then compute the posterior probability of the correct description using a softmax function:
\begin{equation}
p(d^+ \mid A_t) = 
\frac{
    \exp\left( s_t(d^+) \right)
}{
    \exp\left( s_t(d^+) \right) + 
    \sum_{d^{-} \in \mathcal{D}^{-}} \exp\left( s_t(d^{-}) \right)
}
\end{equation}

Note that the posterior $p$ is specific to a batch sample $i$, region $j$, and time frame $t$, from which $A_t$, $d^+$, and $\mathcal{D}^{-}$ are derived as described above. Therefore, the final training objective is the average negative log-likelihood over all region-aligned frames, averaged across all annotated regions $\mathcal{R}$ in the batch:
\begin{equation}
\mathcal{L}_{\text{frame}} = - \frac{1}{|\mathcal{R}|} \sum_{r \in \mathcal{R}} \frac{1}{t_{\text{off}} - t_{\text{on}}} \sum_{t = t_{\text{on}}}^{t_{\text{off}}} \log p(d^+ \mid A_t)
\end{equation}
where we normalize by segment length to ensure equal contribution from each region, regardless of duration.

\section{Experiments}
\label{sec:experiments}
The goal of our experiments is to train sound event detection systems that detect and temporally localize arbitrary sounds based on textual descriptions. To this end, we first pretrain our models on the Clotho dataset~\cite{Drossos20Clotho} with traditional contrastive training on weak caption to improve downstream task performance. We then finetune these models on TACOS using strong captions and the frame-wise loss (Section~\ref{sec:method}).

\subsection{Datasets}

Pretraining is done on the 3,840 audios from Clotho’s train split \cite{Drossos20Clotho} using a contrastive text-audio loss \cite{Elizalde23CLAP}. Each 15–30 second audio recording has five weak human-written captions (8–20 words). After pretraining, the text-audio retrieval performance is evaluated on the 1,045 Clotho test recordings. For fine-tuning, we use the 10,358 strongly annotated audio files from TACOS' train split, excluding any overlapping files with Clotho’s test or validation sets.

\subsection{Audio \& Text Embedding}

We employ RoBERTa \cite{Yinhan19Roberta} to encode textual descriptions into embedding vectors. RoBERTa is a bi-directional transformer-based encoder pretrained via self-supervised learning on the BookCorpus \cite{Zhu15bookcorpus} and WikiText \cite{Stephen17wikitext} datasets. We extract the representation of the [CLS] token from RoBERTa's output as the sentence embedding and project it into the shared audio-text embedding space ($D=1024$) using a linear transformation followed by $\ell^2$-normalization.

For audio, we use a variant of the Audio Spectrogram vIsion Transformer (ASiT) \cite{Ahmed24ASiT}, which was pretrained with group-masked modeling and self-distillation objectives and additionally finetuned on AudioSet weak \cite{Schmid25Pretraining}. ASiT is well suited for our task, as its masked spectrogram modeling encourages the learning of fine-grained local representations.
% \footnote{available here: https://github.com/fschmid56/PretrainedSED}.
To manage computational and memory demands, we divide each audio recording into 10-second segments, embed them to a sequence of frame-wise embeddings independently, and then recombine them using a BiGRU with 768 hidden units. Similar to the text embeddings, we project each embedding vector in the sequence of frame-wise embedding vectors into the $D$-dimensional shared audio-text embedding space using a learned linear transformation followed by $\ell^2$-normalization. The resulting sequence has a maximum length of $T=1491$, corresponding to a temporal resolution of $\delta \approx 20$ms. 
%TODO: aggregation for gloabal contrastive loss.

\subsection{Training}
Pretraining and finetuning use gradient descent with a batch size of 32 and the Adam optimizer \cite{Diederik15adam}. Global contrastive loss is minimized during pretraining; the proposed frame-wise contrastive loss is used for finetuning. During pretraining, modality encoders are jointly optimized over 20 epochs, starting with one epoch of linear warmup. The learning rate increases linearly to $6 \times 10^{-5}$, then decays to $1 \times 10^{-7}$ via cosine annealing. The temperature $\tau$ is set to 0.05, based on Clotho validation performance. Finetuning on TACOS uses the same setup, except with 6 epochs and a higher peak learning rate of $6 \times 10^{-4}$. Hyperparameter search over temperature values $\tau \in \{0.01, 0.05, 0.1, 0.2, 0.3, 0.4\}$ is conducted to determine the optimal setting.

\subsection{Evaluation Setup}
\label{sec:eval_setup}

To identify regions where a sound event is active, we embed both the class description and the audio recording into a shared audio-text embedding space using the modality encoders $f_d$ and $f_a$, respectively. We then compute the cosine similarity between the embedded text and the sequence of audio embedding vectors, yielding a sequence of detection scores. 
We evaluate the resulting predictions on the AudioSet benchmark with temporally strong annotations~\cite{Hershey21StrongLabels}. Specifically, we select 117 of the 416 original AudioSet categories and map them to 50 of the 59 subclasses in the Tacos taxonomy (e.g., \textit{'Insect', 'Bee, wasp, etc.', 'Mosquito', 'Fly, housefly'} $\rightarrow$ \textit{'Insect Buzz'}). The remaining 9 classes (mostly music instruments) are excluded due to missing strong annotations in AudioSet. Files that do not contain any annotations are excluded. For each of the 50 selected categories, we construct a single-sentence textual description with \textit{gpt-4o-mini-2024-07-18} (e.g., \textit{'Insect Buzz'} $\rightarrow$ \textit{'Insects are buzzing.'}), which serves as input to the text encoder during evaluation. In the results section, we report the segment-level pAUROC metric with a maximum false positive rate of 0.1 using 1-second segments, as well as the event-level PSDS1 score \cite{Bilen20PSDS}, \cite{Ebbers22PSDS_} without applying the variance penalty.

\section{Results and Discussion}
\label{sec:results}

Table~\ref{tab:clotho} reports the retrieval performance of our model after pretraining on the Clotho benchmark. The results are comparable to those achieved by similar dual encoder models proposed in \cite{Wu23LION_CLAP} and \cite{Primus24Retrieval}.
%, which are trained on multiple datasets and utilize a larger text encoder, respectively. 
To assess the usefulness of weak captions generated by TACOS, we merge them with Clotho for pretraining and observe an improvement in text-audio retrieval performance of approximately 1.2 percentage points in mAP@10. This suggests that the LLM-summarized weak captions in TACOS are beneficial for contrastive audio-language pretraining.

\begin{table}[h]
\centering
\caption{Text-to-Audio retrieval performance on the Clotho benchmark.}
\label{tab:clotho}
\begin{tabular}{@{}lcccc@{}}
\toprule
\textbf{Method} & \textbf{mAP@10} & \textbf{R@1} & \textbf{R@5} & \textbf{R@10} \\ \midrule
LION-CLAP \cite{Wu23LION_CLAP}         & --     & 17.20 & 42.90 & 55.40 \\
Primus et al. \cite{Primus24Retrieval}     & 28.93  & 18.11 & 43.54 & 57.57 \\
Ours (pretraining)                      & 29.29  & 18.36 & 43.92 & 57.65 \\
\quad + Tacos-Weak        & \textbf{30.53}  & \textbf{19.16} & \textbf{45.57} & \textbf{60.27} \\ \bottomrule
\end{tabular}
\end{table}

%We evaluate the sound event detection performance of our proposed method as outlined in Section~\ref{sec:experiments}, with results  Specifically, 
We compare our method---fine-tuning with strong captions and the frame-wise loss---against the models after pretraining on Clotho and the second variant that was fine-tuned using only TACOS' weak captions with standard contrastive audio-text training. The results are presented in Table~\ref{tab:audioset_selected}.
%For both fine-tuning settings, the table also reports the optimal temperature parameter $\tau$.

Weak fine-tuning yields improvements of 7.8 pp. PSDS1 and 6.35 pp. pAUROC over the pretrained model. 
Using the strong captions for fine-tuning leads to a PSDS1 gain of 13.3 pp. and a pAUROC gain of 11.9 pp. compared to the pretrained model; this corresponds to a PSDS1 and pAUROC improvement of 5.5 pp. and 5.61 pp., respectively, compared to the weakly fine-tuned model. 

Table~\ref{tab:audioset_selected} further shows the detection performance of an ASiT model trained in a fully supervised manner using the strong annotations from AudioSet \cite{Schmid25Pretraining}. To compare, we mapped the 117 AudioSet classes to the 50 subcategories in TACOS, as described in Section~\ref{sec:eval_setup}. The supervised baseline outperforms our text-audio-based detection system by 36.27 pp. in PSDS1 and 8.97 pp. in pAUROC. We attribute this substantial gap to the additional strongly annotated training data, as well as the inherent challenge of matching free-text descriptions to audio events, rather than detecting a fixed set of predefined categories.

To further strengthen the validity of our results, we also report the detection performance on all 416 classes in AudioSet Strong in the second section of Table~\ref{tab:audioset_selected}. Instead of creating a caption-style single sentence descriptions for each category, we use the keywords-style class names as class descriptor. The results reveal a consistent ranking among the pretrained, weakly finetuned, and strongly finetuned models across both evaluation metrics. However, overall detection performance declines noticeably compared to the evaluation on TACOS subclasses, highlighting the inherent difficulty of zero-shot generalization to concepts that are underrepresented in the training set.

\begin{table}[ht]
\centering
\caption{Sound event detection performance on AudioSet. The top section reports results on 117 AudioSet classes mapped to sound sub categories in TACOS. The bottom section shows results on all 416 AudioSet classes.}
\label{tab:audioset_selected}
\begin{tabular}{@{}lccc@{}}
\toprule
\textbf{Method} & \textbf{PSDS1} & \textbf{pAUROC} & \boldmath{$\tau$} \\
\midrule
\multicolumn{4}{l}{\textit{Evaluation on 50 sound categories in TACOS}} \\
pretraining            & \;\;4.61 $\pm$ 0.23 & 66.74 $\pm$ 0.30 & -   \\
--- weak finetuning     & 12.44 $\pm$ 0.06    & 73.09 $\pm$ 0.20 & 0.2 \\
--- strong finetuning   & 17.99 $\pm$ 0.10    & 78.70 $\pm$ 0.10 & 0.1 \\
supervised training \cite{Schmid25Pretraining} &                     54.26\;\;\;\;\;\;\;\;\;\;\;\; &      87.67\;\;\;\;\;\;\;\;\;\;\;\;            &     \\ 
\midrule
\multicolumn{4}{l}{\textit{Evaluation on all 416 AudioSet classes}} \\
pretraining            &  1.17 $\pm$ 0.03    & 39.19 $\pm$ 0.01 & -   \\
--- weak finetuning     &  3.21 $\pm$ 0.01    & 43.27 $\pm$ 0.02 & 0.2 \\
--- strong finetuning   &  6.24 $\pm$ 0.01    & 50.11 $\pm$ 0.01 & 0.1 \\
\bottomrule
\end{tabular}
\end{table}

\begin{figure}[!t]
\centering
\includegraphics[width=0.8\linewidth]{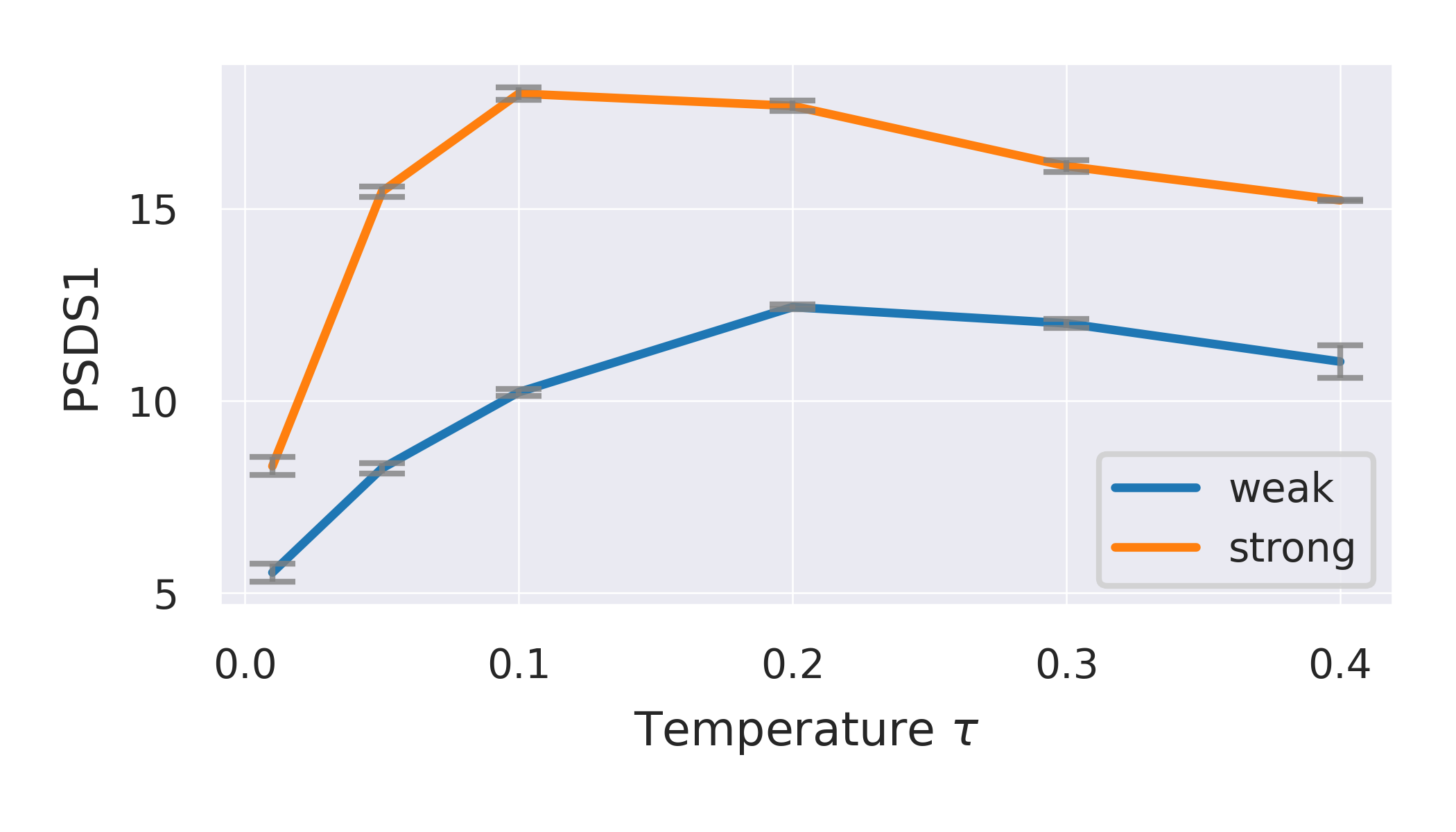}
\vspace{-0.5cm}
\caption{PSDS1 scores across a range of temperature values ($\tau$) for models trained with weak and strong annotations in TACOS.}
\label{fig:experiments_tau}
\end{figure}

To provide a more comprehensive view of the model's performance across different values of the temperature parameter $\tau$, we plot the PSDS1 metric over a range of $\tau$ values in Figure~\ref{fig:experiments_tau}. The slight differences in the optimal modes suggest that the weakly and strongly fine-tuned variants benefit from independent selection of $\tau$ to achieve optimal detection performance. Overall, the model fine-tuned with the strong annotations and the frame-wise loss achieves better 
% detection 
performance compared to the model fine-tuned on the weak annotations.

\section{Conclusion}
\label{sec:conclusion}

In this paper, we introduced TACOS, a new dataset consisting of 12,358 audio recordings with 47,748 temporally-aligned free-text captions. We used large language models to refine the captions and generate additional weak captions from the strong ones. Building on this resource, we extended the contrastive text-audio training commonly used in CLAP-style models to a frame-wise variant, exploiting the strong captions. The results show that the introduced contrastive learning strategy applied to the strong captions of TACOS substantially improves the performance on the task of free text-based Sound Event Detection on AudioSet Strong compared to training only on weak captions. Beyond audio retrieval and text-based sound event detection, we hope that temporally-precise captions will also enable progress in tasks such as text-conditioned audio generation, audio captioning, and audio question answering.

\bibliographystyle{IEEEtran}
\bibliography{refs25}

\end{document}